\documentclass[10pt, conference]{IEEEtran}
\IEEEoverridecommandlockouts

\usepackage{cite}
\usepackage{amsmath,amssymb,amsfonts}
\usepackage{algorithmic}
\usepackage{graphicx}
\usepackage{textcomp}
\usepackage[table]{xcolor}
\usepackage{braket}

\usepackage{comment}
\usepackage{booktabs}
\usepackage{here}
\usepackage{subfigure}
\usepackage{physics}
\usepackage{ascmac}
\usepackage[hyphens]{url}
\usepackage{multirow}
\usepackage{hyperref}
\usepackage{xspace}
\usepackage{framed}
\usepackage{inconsolata}
\usepackage{threeparttable}
\usepackage{pifont}
\usepackage{circledsteps}
\usepackage{listings}
\usepackage{tcolorbox}
\usepackage{enumitem}
\usepackage{colortbl}
\usepackage{bm}
\usepackage{stfloats}
\usepackage{xcolor}

\usepackage{float}
\def\fig#1{Figure \ref{#1}}
\def\tab#1{Table \ref{#1}}

\def\eg{e.g.\xspace}
\def\ie{i.e.\xspace}

\definecolor{diffstart}{named}{gray}
\definecolor{diffincl}{rgb}{0.0, 0.5, 0.0} 
\definecolor{diffrem}{named}{purple}

\lstdefinelanguage{diff}{
language=python,
basicstyle=\ttfamily\small,
columns=fullflexible,
float,
floatplacement=t,
frame=tb,
morecomment=[f][\color{diffrem}]{-}, 
morecomment=[f][\color{diffincl}]{+}, 
}

\def\summaryblock#1#2{
\begin{oframed}
\noindent \textbf{#1:}
#2
\end{oframed}
}

\newcommand{\astagen}{\textsc{AstaGen}\xspace}

\newcommand{\red}[1]{\textcolor{black}{#1}}

\begin{document}

\title{How Far Have LLMs Come Toward Automated SATD Taxonomy Construction?}


\author{
\IEEEauthorblockN{
   Sota Nakashima\IEEEauthorrefmark{1},
   Yuta Ishimoto\IEEEauthorrefmark{1},
   Masanari Kondo\IEEEauthorrefmark{1},
   Tao Xiao\IEEEauthorrefmark{1},
   Yasutaka Kamei\IEEEauthorrefmark{1},
}
\IEEEauthorblockA{
   \IEEEauthorrefmark{1}Kyushu University, Fukuoka, Japan \\
   nakashima@posl.ait.kyushu-u.ac.jp, ishimoto@posl.ait.kyushu-u.ac.jp, \\
   kondo@ait.kyushu-u.ac.jp, xiao@ait.kyushu-u.ac.jp, kamei@ait.kyushu-u.ac.jp
}
}


\maketitle

\begin{abstract}
Technical debt refers to suboptimal code that degrades software quality. When developers intentionally introduce such debt, it is called \emph{self-admitted technical debt (SATD)}.
Since SATD hinders maintenance, identifying its categories is key to uncovering quality issues.
Traditionally, constructing such taxonomies requires manually inspecting SATD comments and surrounding code, which is time-consuming, labor-intensive, and often inconsistent due to annotator subjectivity.
In this study, we investigated to what extent large language models (LLMs) could generate SATD taxonomies. We designed a structured, LLM-driven pipeline that mirrors the taxonomy construction steps researchers typically follow.
We evaluated it on SATD datasets from three domains: quantum software, smart contracts, and machine learning.
It successfully recovered domain-specific categories reported in prior work, such as \textit{Layer Configuration} in machine learning.
It also completed taxonomy generation in under two hours and for less than \$1, even on the largest dataset.
These results suggest that, while full automation remains challenging, LLMs can support semi-automated SATD taxonomy construction.
Furthermore, our work opens up avenues for future work, such as automated taxonomy generation in other areas.

\end{abstract}


\vspace{-1mm}
\section{Introduction} \label{intro}
\vspace{-1mm}

Developers sometimes introduce suboptimal code that degrades software quality; this is known as \emph{technical debt}~\cite{cunningham1992wycash}. In particular, when developers intentionally insert such code due to constraints such as budget or schedule, it is referred to as \emph{Self-Admitted Technical Debt (SATD)}~\cite{potdar2014exploratory}. In practice, developers explicitly mark SATD via code comments.

Because SATD can negatively affect software maintenance, identifying its categories can improve code quality~\cite{maldonado2015mtd,bavota2016large}.
Previous studies therefore developed reusable SATD taxonomies that systematically organize these debts~\cite{maldonado2015mtd,ishimoto2024quantum,ebrahimi2023blockchain}.
While these taxonomies transfer within a domain (e.g., Java), new domains (e.g., quantum software) require fresh ones.



Although SATD taxonomies help debt detection and remediation, constructing them requires significant manual effort, typically involving at least two researchers to analyze comments~\cite{maldonado2015mtd,ishimoto2024quantum,ebrahimi2023blockchain,obrien2022ml}. 
Such manual analysis is often subjective and lacks reproducibility.
As a result, automating SATD taxonomy construction remains a critical challenge.

In this study, we empirically investigated to what extent large language models (LLMs) could generate SATD taxonomies. To that end, we designed a structured pipeline that first prompts an LLM to generate concise explanations for each SATD comment and surrounding source code, then iteratively proposes and refines categories based on those explanations, reproducing the actual manual steps that researchers follow when building SATD taxonomies~\cite{bavota2016large, ishimoto2024quantum, xiao2021characterizing}.

We evaluated this pipeline across three domains (\ie, quantum software~\cite{ishimoto2024quantum}, smart contracts~\cite{ebrahimi2023blockchain}, and machine learning software~\cite{obrien2022ml}) using human-defined taxonomies as references.
It successfully generated domain-specific categories that were semantically similar to those defined by humans, such as \textit{Layer Configuration} in machine learning,
yielded more consistent taxonomy compared to a naive use of an LLM, and completed taxonomy generation in under two hours and at a cost of less than \$1, even for the largest dataset (448 comments), highlighting its efficiency compared to manual approaches.

Our main contributions are as follows:
(1) We investigated to what extent LLMs could generate SATD taxonomies.
(2) We conducted an extensive evaluation across three domains.
(3) We suggested practical use cases in semi-automated SATD taxonomy generation supported by our empirical results.

\section{Related Work} \label{RelatedWork}

\begin{figure}[b]
    \centering
    \includegraphics[width=0.65\linewidth]{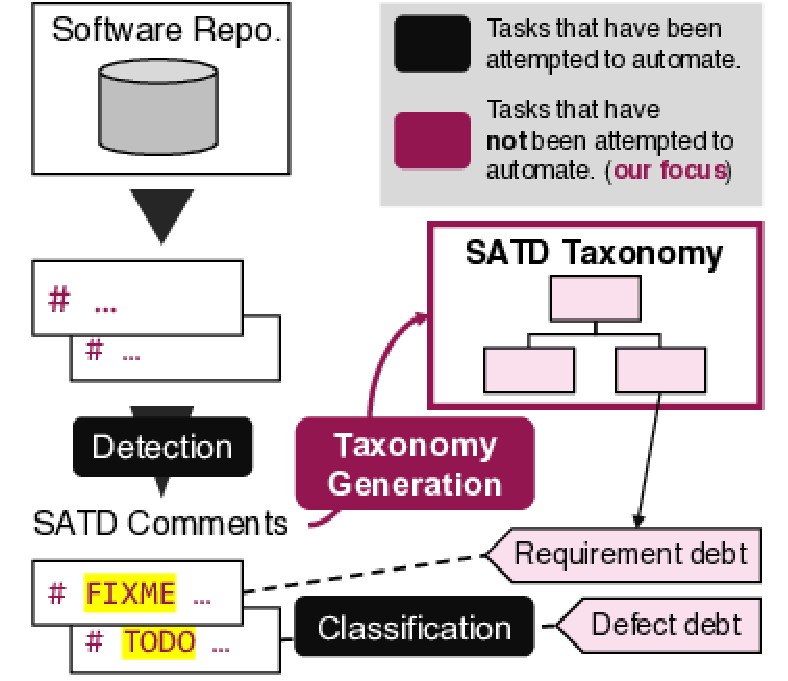}
    \caption{SATD detection, classification, and taxonomy generation.}
    \label{fig:satd_flow}
\end{figure}

\vspace{-1mm}

\subsection{SATD Taxonomy and Automation} \label{sec:satd-background}

To understand technical debt, prior work has analyzed the types and frequencies of SATD~\cite{maldonado2015mtd,bavota2016large,ishimoto2024quantum,ebrahimi2023blockchain,xiao2021characterizing}. Constructing a taxonomy is key: researchers first filter comments using keyword patterns (e.g., \texttt{todo}, \texttt{fixme})~\cite{potdar2014exploratory,maldonado2015mtd}, then two or more reviewers propose and reconcile category labels until consensus. While essential, this process is time-consuming (120 person-hours for 500 comments~\cite{xiao2021characterizing}), unscalable (375 of 15,671 comments annotated~\cite{kashiwa2022ist}), and prone to bias (12 disagreements in 50 comments~\cite{ebrahimi2023blockchain}), motivating automation.

Automation efforts have focused on SATD detection and classification (Fig.~\ref{fig:satd_flow}). Maldonado et al.~\cite{maldonado2017tse} and Yu et al.~\cite{yu2020tse} applied traditional ML to detect SATD, while Chen et al.~\cite{chen2021tr} used XGBoost to classify three types (defect, design, implementation). More recently, Sheikhaei et al.~\cite{sheikhaei2024emse} fine-tuned and prompted LLMs to classify SATD into predefined categories with improved accuracy. However, none addressed taxonomy generation itself. Our work fills this gap by exploring LLM-driven construction of new SATD category hierarchies.

\vspace{-2mm}

\subsection{Taxonomy Construction.}
Inducing taxonomies from unstructured data, such as code comments or social media posts, is challenging due to domain heterogeneity and the absence of a fixed structure. Durham et al.~\cite{durham2023unveiling} used topic modeling to derive category schemas from tweets, and Najem et al.~\cite{najem2021semi} proposed semi-automatic ontology building via semantic networks. 

Chen et al. \cite{chen2020constructing} leveraged pretrained transformer models to assemble taxonomic hierarchies, and compared prompting and fine‐tuning strategies for hypernym extraction \cite{chen2023prompting}. 
More recently, Shah et al. \cite{shah2023using} and Wan et al. \cite{wan2024tnt} proposed end‐to‐end LLM pipelines that generate, refine, and assign category labels from unstructured natural language for user‐intent analysis in log data. 
We adopted this paradigm, but extended to SATD taxonomy by jointly considering each comment and its surrounding code.

\begin{figure*}[t]
    \centering
    \includegraphics[width=0.8\linewidth]{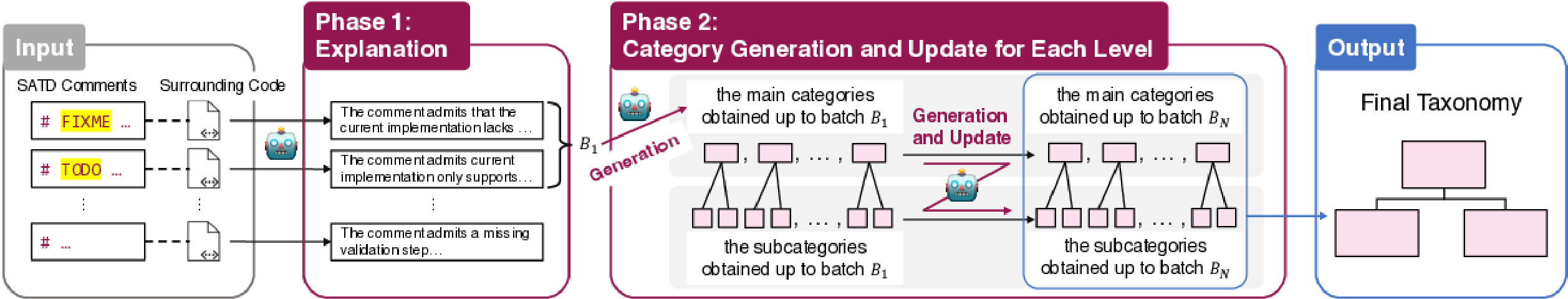}
    \caption{The LLM-driven pipeline for SATD taxonomy generation.
    The robot icons above the arrows indicate the steps in which LLMs are utilized.}
    \label{fig:workflow}
\end{figure*}

\vspace{-1mm}
\section{LLM-driven pipeline \\for SATD taxonomy generation} \label{sec:methodology}
\vspace{-1mm}

\subsection{Overview}
Figure~\ref{fig:workflow} illustrates our overall workflow. Since taxonomy generation comprises topic extraction and hierarchical organization, we adopt a multi-phase pipeline. Similar multi-phase approaches have proven effective in other domains~\cite{wan2024tnt, shah2023using}.

\noindent \textbf{Input and Output.}
The input consists of SATD comments and their corresponding source code.  
The output is a SATD taxonomy generated by the pipeline.
The resulting taxonomy is structured as a two-level tree.  
We refer to the upper and lower levels of this hierarchy as the \textit{main categories} and \textit{subcategories}, respectively.
This structure aligns with most existing studies on SATD taxonomy construction~\cite{bavota2016large, ebrahimi2023blockchain, kashiwa2022ist, ishimoto2024quantum}, which adopt trees with at least two hierarchical levels.



\noindent \textbf{Workflow.} Our process mirrors human SATD taxonomy construction by (1) generating concise explanations for each SATD comment in its code context and then (2) iteratively proposing, merging, and refining category labels~\cite{bavota2016large, ishimoto2024quantum, xiao2021characterizing}. 


\subsection{Phase 1: Generating Explanations for SATD} \label{sec:phase1}


In this phase, the LLM generates a concise explanation (typically 2--3 sentences) for each SATD comment. To capture both the comment and its related code context, we feed the entire source file as context, since prior work has shown that surrounding code aids SATD interpretation~\cite{sheikhaei2024emse,yonekura2025icpc}.



\subsection{Phase 2: Generating/Updating Categories at Each Level} \label{sec:phase2}


The LLM builds the SATD taxonomy hierarchically via iterative batches. First, all main categories are generated, followed by the subcategories associated with each.
For each batch of explanations, the LLM proposes new labels and merges them into the existing list until all comments are covered. 
Below are the generation and update steps.


\subsubsection{Generation Step} \label{sec:generation}

In this step, the LLM generates a category name of approximately 1--3 words for each comment included in the batch, based on its corresponding explanation.
All explanations in the current batch are enumerated and presented together in a single prompt.  
The LLM then produces a category name for each explanation individually.

\subsubsection{Update Step} \label{sec:update}

In the update step, new categories from the current batch are merged with those accumulated so far. The LLM receives the combined list of category names and their explanations and is prompted to merge semantically similar labels. By comparing the lists before and after merging, we derive a mapping of renamed categories and apply it to the master list; no further LLM calls are needed. Finally, each comment is assigned to its category within the resulting two-level taxonomy of main and subcategories.

\vspace{-1mm}
\section{Study Design}\label{setting}
\vspace{-1mm}

\subsection{Research Questions} \label{rq}
\newcommand{\RQone}{\textbf{(Category name alignment)} Can the pipeline generate category names that are semantically similar to those defined by humans?}

\newcommand{\RQtwo}{\textbf{(Taxonomy content alignment)} \red{Does the pipeline achieve better alignment with human-defined taxonomies than a naive use of an LLM, in terms of comment assignment and category granularity?}}
\newcommand{\RQthree}{\textbf{(Efficiency)} How much time and cost does the pipeline require?}

\begin{enumerate}[label=\textit{RQ\arabic*}, leftmargin=*, align=left]
    \item \RQone
    \item \RQtwo
    \item \RQthree
\end{enumerate}
RQ1 focuses on the similarity of category names, while RQ2 examines the consistency of the comments assigned to each category.
RQ3 assesses the practicality of automated SATD taxonomy generation by measuring the execution time and monetary cost (\ie, LLM API fees).

\subsection{Studied Datasets}
We used SATD comments and corresponding taxonomies from prior studies as our datasets.
The SATD comments are input to LLMs, while the human-defined taxonomies serve as references for evaluation.

We collected papers that constructed SATD taxonomies from two major digital libraries: IEEE Xplore and the ACM Digital Library.
Specifically, we used the following query to search for papers whose abstracts contained specific keywords:
\texttt{("SATD" OR "self-admitted technical debt") AND ("taxonomy" OR "category" OR "coding")}.  
After removing duplicates, this query returned 79 papers.\footnote{
Search conducted on January 8, 2025; reconfirmed unchanged on May 16, 2025. 
The full list is available in the replication package.}

From these, we selected papers that met both of the following criteria:
(1) the taxonomy was manually constructed, and  
(2) both the replication package and its underlying data were publicly available.  
Two authors independently screened the papers, resulting in three candidate papers~\cite{obrien2022ml,ebrahimi2023blockchain,ishimoto2024quantum}:




\begin{itemize}\setlength{\itemsep}{0pt}
  \item \textbf{Quantum Software (QS)}~\cite{ishimoto2024quantum}: 88 SATD comments from Qiskit\cite{javadi2024qiskit} projects; includes 4 main and 9 subcategories.
  \item \textbf{Smart Contracts (SC)}~\cite{ebrahimi2023blockchain}: 190 SATD comments from Solidity contracts; includes 6 main and 26 subcategories.
  \item \textbf{Machine Learning Software (ML)}~\cite{obrien2022ml}: 448 SATD comments from ML-related GitHub repositories; includes 9 main and 23 subcategories.
\end{itemize}

\subsection{Selected Models}
In all experiments, we used DeepSeek-V3~\cite{liu2024deepseek} with the default temperature setting of 1.0 (recommended for data analysis tasks by the official API documentation\footnote{\url{https://api-docs.deepseek.com/quick_start/parameter_settings}}) and ran each experiment ten times to assess result stability.
We chose Deepseek-V3 for its low per-request cost while retaining competitive inference performance compared to other models~\cite{liu2024deepseek}.

In RQ1, we measured semantic similarity between human-defined and LLM-generated category names by encoding each with the \texttt{all-MiniLM-L6-v2} sentence embedding model\footnote{\url{https://huggingface.co/sentence-transformers/all-MiniLM-L6-v2}} and computing cosine similarity. 
We selected a sentence-level embedding model because category names often consist of multiple words, and such models are better suited for capturing their overall semantics.

\subsection{Prompt Design}
In Phase 1 (Section~\ref{sec:phase1}), we limited the source code context given to the LLM to 2,000 lines to comply with its context length limit.
If the file exceeded this limit, we included only the 1,000 lines before and after the SATD comment.  


In Phase 2 (Section \ref{sec:phase2}), we processed SATD explanations in batches of 20, consistent with the batching strategy in prior work~\cite{ishimoto2024quantum}. All prompts are included in our replication package.\footnote{\url{https://doi.org/10.5281/zenodo.17355443}}

\subsection{Evaluation Metrics} \label{sec:metrics}
This section describes the evaluation metrics used in each RQ.
For each dataset and taxonomy level, let $\mathcal{H}$ and $\mathcal{M}$ be the sets of human-defined and LLM-generated category names, respectively.  
We denote individual category names from these sets as $H_i \in \mathcal{H}$ and $M_j \in \mathcal{M}$.\footnote{$\mathcal{M}$ stands for ``machine-generated,'' in contrast to $\mathcal{H}$ for ``human-defined.''}

\subsubsection{RQ1}
We computed the cosine similarity between the sentence embeddings for $H_i$ and $M_i$, denoted as $sim(H_i, M_j)$.
We then computed $top\_sim(H_i) = \max_j sim(H_i, M_j)$, which represents the highest similarity between $H_i$ and any generated name.
This score quantifies how closely LLMs can generate a category name for each human-defined category.
We visualized the distribution of $top\_sim(H_i)$ for each dataset and level to assess the alignment between human-defined and LLM-generated category names.

\begin{figure}[t]
    \centering
    \includegraphics[width=0.65\linewidth]{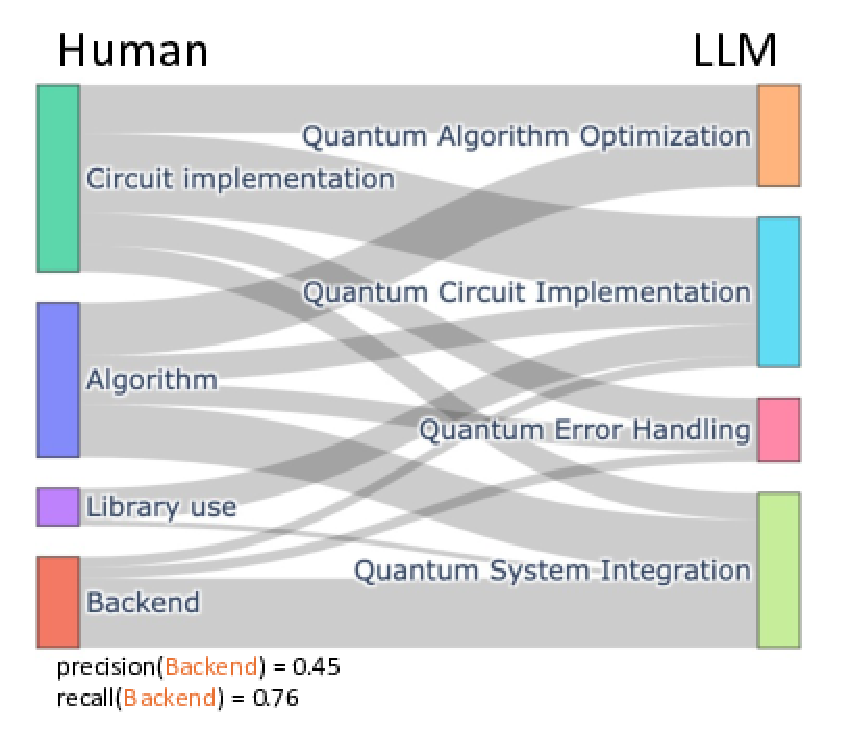}
    \caption{Example Sankey diagram of human-defined and LLM-generated categories (Dataset: QS, Level: Main, Run ID: 0).}
    \label{fig:sankey}
\end{figure}

\subsubsection{RQ2}
We evaluated whether similar comments were grouped into similar categories by humans and LLMs.
Unlike classification problems, $H_i$ and $M_j$ are drawn from different sets (\ie, $\mathcal{H}$ and $\mathcal{M}$), making conventional precision and recall inapplicable.
To address this, we define a \textit{best-match precision and recall} for each human-defined category.
Let $C_{ij}$ denote the number of comments assigned to both $H_i$ (by humans) and $M_j$ (by LLMs).
The matrix $C$ enables the construction of a Sankey diagram (\eg, \fig{fig:sankey}) that visualizes the flow of comments between human-defined and LLM-generated categories.\footnote{The thickness of each flow from $H_i$ to $M_j$ is proportional to $C_{ij}$.}
For each $H_i$, we define its \textit{best-matched category} as $M_{j^*}$, where $j^* = \operatorname{argmax}_j C_{ij}$.
Using this best match, we compute:
\begin{align}
    \text{precision}(H_i) = \frac{C_{ij^*}}{\sum_{i=1}^{|\mathcal{H}|} C_{ij^*}}, \quad
    \text{recall}(H_i)    = \frac{C_{ij^*}}{\sum_{j=1}^{|\mathcal{M}|} C_{ij}}. \nonumber
\end{align}
Here, precision reflects how exclusively the best-matched category corresponds to the human-defined category $H_i$, while recall indicates how well the comments in $H_i$ are covered by the best-matched category.
We also report the F1 score as the harmonic mean of precision and recall.

Our metrics quantify category correspondence via the Sankey diagram. For example, Fig.~\ref{fig:sankey} shows human-defined “Backend” best matching LLM’s “Quantum System Integration”: most “Backend” comments fall under that LLM category (high recall), but that category also covers other human-defined labels (lower precision). 
Because recall can be inflated when the LLM uses fewer categories, we also introduce \textit{category granularity}, defined as \(|\mathcal{M}|-|\mathcal{H}|\). Values closer to zero indicate that the LLM’s category count matches the human-defined count more closely.

\subsubsection{RQ3}
We measure the average runtime and DeepSeek-V3 API cost over ten runs per dataset. Each run logs total input/output tokens, and cost is computed using the official DeepSeek-V3 pricing table.\footnote{\url{https://api-docs.deepseek.com/quick_start/pricing}, as of May 20, 2025}

\vspace{-1mm}
\section{Results} \label{result}
\vspace{-1mm}
\subsection{RQ1: Category Name Alignment}

\begin{figure}[t]
    \centering
    \includegraphics[width=0.9\linewidth]{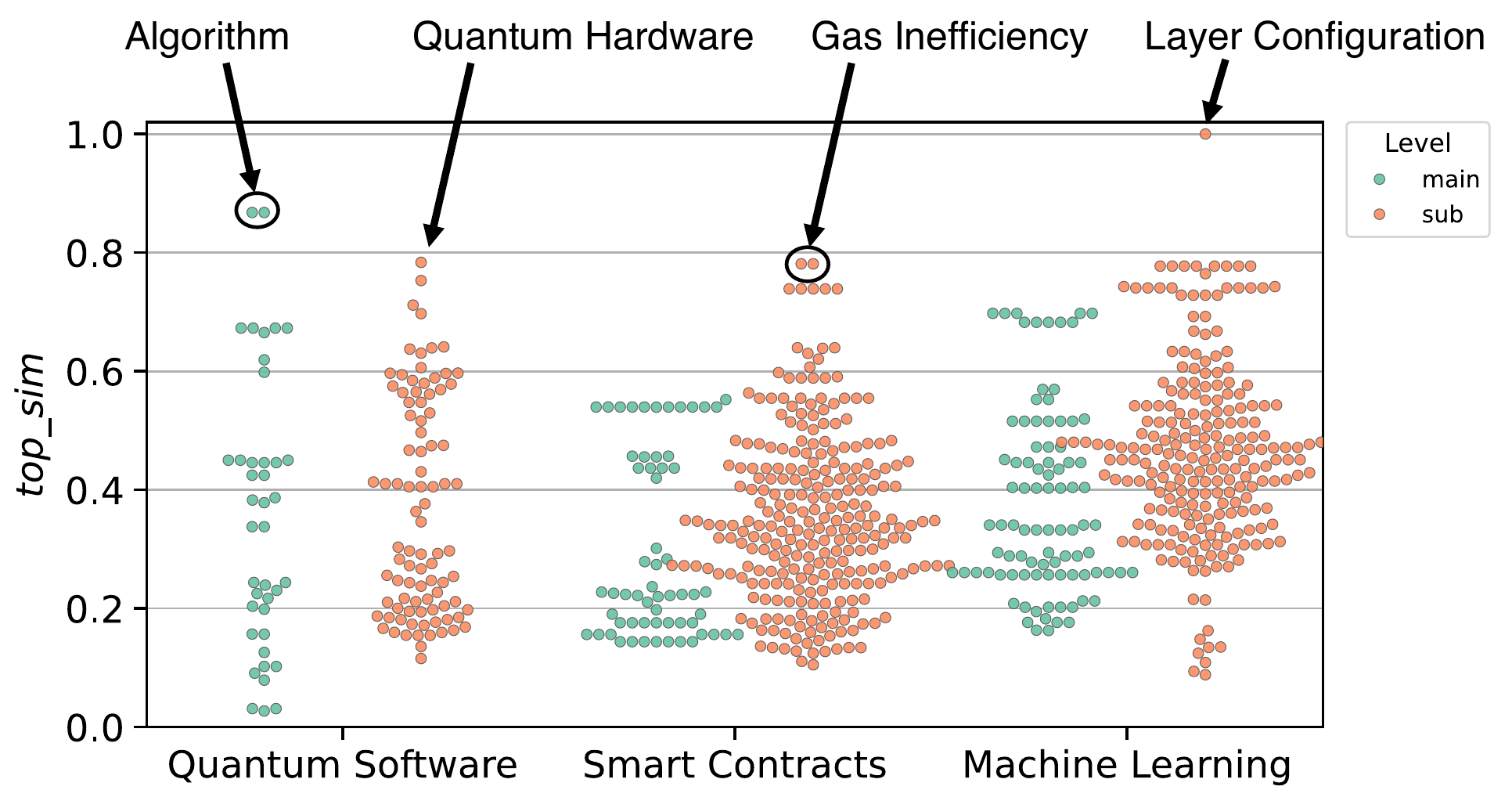}
    \caption{
    Distribution of $top\_sim(H_i)$ scores across datasets and taxonomy levels. 
    The annotated examples (e.g., \textit{Algorithm}, \textit{Layer Configuration}) correspond to the top-5 highest similarity scores across all data points.
    }
    \label{fig:rq1_swarm}
\end{figure}

%
\fig{fig:rq1_swarm} shows the distribution of $top\_sim(H_i)$ scores.
The x-axis denotes the dataset, and colors indicate the taxonomy level.
The mean $top\_sim$ scores for the QS, SC, and ML datasets were 0.36, 0.32, and 0.42, respectively.

\textbf{LLMs can generate domain-specific category names that are similar to human-defined ones.}
As highlighted in \fig{fig:rq1_swarm}, all points with scores $\ge 0.8$ corresponded to domain-specific categories (\eg, \textit{Layer Configuration} in ML).

\textbf{High-similarity categories are more likely to be domain-specific, whereas low-similarity categories tend to be generic.}  
To examine the relationship between $top\_sim(H_i)$ and the type of category, two authors independently labeled each human-defined category in the top and bottom 10\% of $top\_sim(H_i)$ as either \textit{domain-specific} or \textit{generic}, counting only those with agreement. 
Of the top 10\% (77 categories), 46 (60\%) were domain-specific; in the bottom 10\%, 53 (69\%) were generic.
These results suggest that LLMs are more effective at generating consistent names for domain-specific categories, while generic categories are harder to match semantically.





\vspace{-2mm}
\summaryblock{Answer to RQ1}{
The pipeline can generate domain-specific category names that are semantically similar to human-defined names.
In contrast, low-similarity cases were predominantly generic (69\% of 77 categories).
}
\vspace{-2mm}

\vspace{-1mm}
\subsection{RQ2: Taxonomy Content Alignment}
\vspace{-1mm}
\begin{table}[t]
\centering
\caption{Comparison of the pipeline and its naive variant.}
\label{tab:pre_rec}
\small
\scalebox{0.70}{
\begin{tabular}{llrr|rr|rr|rr}
\toprule
Data. & Lv.
& \multicolumn{2}{c}{Pre.}
& \multicolumn{2}{c}{Rec.}
& \multicolumn{2}{c}{F1} 
& \multicolumn{2}{c}{$|\mathcal{M}|-|\mathcal{H}|$} \\
& & P & N & P  & N & P & N & P & N \\
\midrule
\multirow{2}{*}{QS} 
    & Main & \textbf{0.43} & 0.25 & 0.59 & \textbf{0.77} & \textbf{0.50} & 0.37 & \textbf{0.3} & 0.5 \\
    & Sub  & \textbf{0.33} & 0.16 & 0.39 & \textbf{0.68} & \textbf{0.36} & 0.26 & 1.9 & \textbf{1.5}\\
\midrule
\multirow{2}{*}{SC} 
    & Main & 0.20 & \textbf{0.27} & 0.63 & \textbf{0.80} & 0.31 & \textbf{0.41} & -2.0 & \textbf{-1.6} \\
    & Sub  & \textbf{0.19} & 0.14 & 0.56 & \textbf{0.79} & \textbf{0.29} & 0.25 & \textbf{-8.2} & -15.5\\
\midrule
\multirow{2}{*}{ML} 
    & Main & \textbf{0.15} & 0.12 & 0.45 & \textbf{0.87} & \textbf{0.22} & 0.20 & \textbf{1.6} & -6.4\\
    & Sub  & \textbf{0.17} & 0.11 & 0.36 & \textbf{0.67} & \textbf{0.23} & 0.19 & \textbf{6.1} & -13.5\\
\bottomrule
\multicolumn{10}{l}{P: The pipeline, N: Naive. Bold indicates the better value.}
\end{tabular}
}
\end{table}

\tab{tab:pre_rec} presents the mean precision, recall, F1 scores, and category granularity for each dataset and taxonomy level.
For comparison, we introduced a baseline, the \textit{naive variant}, which generates the taxonomy directly using an LLM without Phase 1 and 2.
This baseline allows us to evaluate the extent to which the pipeline improves alignment with the human-defined taxonomy compared to a naive use of LLMs.

\textbf{The pipeline achieved higher F1 scores than the naive variant in five out of six cases, indicating a better balance between precision and recall.}  
This result suggests that core components, which the naive variant lacks, are effective in improving the alignment of comment assignments.
\textbf{It also matched the granularity of human-defined taxonomies more closely in four of six cases,} exhibiting a smaller absolute category-count difference than the naive variant.
In contrast, the naive variant often produced fewer categories, which likely contributed to its higher recall.



\vspace{-2mm}
\summaryblock{Answer to RQ2}{
The pipeline outperformed the baseline in F1 (5/6 cases) and generated better category counts (4/6 cases). Precision and recall still lay between 0.152 and 0.638, showing room for improvement.
}
\vspace{-2mm}

\begin{table}[t]
\centering
\caption{Average execution time and cost for each dataset.}
\label{tab:efficiency}
\small
\scalebox{0.75}{ 
\begin{tabular}{lrr}
\toprule
Dataset (Size) & Avg. Time [min] & Avg. Cost [USD] \\
\midrule
QS (88)   & 21.60 & 0.142 \\
SC (190)  & 52.98 & 0.614 \\
ML (448)  & 109.50 & 0.813 \\
\bottomrule
\end{tabular}
}
\end{table}

\vspace{-2mm}
\subsection{RQ3: Efficiency (Time and Cost)}
\vspace{-1mm}

\tab{tab:efficiency} shows the average execution time and cost for each dataset, averaged over ten runs.

\textbf{Compared to prior manual efforts, LLMs achieved significantly reduces both time and cost.}
For example, Ebrahimi et al.~\cite{ebrahimi2023blockchain} reported spending 57 person-hours just to construct the initial taxonomy, while Xiao et al.~\cite{xiao2021characterizing} required 120 person-hours to manually inspect 500 SATD comments.
These substantial efficiency gains highlight the practicality of using LLMs for large scale SATD taxonomy generation.


\vspace{-2mm}
\summaryblock{Answer to RQ3}{
Laverage LLMs in SATD taxonomy construction is efficient in terms of both time and cost.
Even for the largest dataset (\ie, 448 comments), it took only 109.50 minutes and cost just \$0.813 on average.
}
\vspace{-2mm}

\vspace{-2mm}
\section{Use Case} \label{usecase}
\vspace{-2mm}

While our results indicate that LLMs cannot perfectly replicate human-defined SATD taxonomies, they can meaningfully reduce manual effort. We propose two use cases: (1) as a virtual annotator that can replace some human annotators in collaborative taxonomy construction (e.g., standardizing coding guides) \cite{treude2025harmonized} and (2) as an initial step that provides category candidates to support annotators.
Notably, LLMs demonstrated the ability to generate domain-specific categories. This makes it especially valuable when researchers must construct a taxonomy in an unfamiliar domain.

\vspace{-1mm}
\section{Threats to Validity} \label{sec:threats}
\vspace{-1mm}



\noindent \textbf{Construct Validity.}
Our RQ1 and RQ2 evaluations measured only the alignment with the human-defined taxonomy.
In RQ1, some LLM-generated categories that did not align could represent novel category discoveries.
Future work will investigate whether such categories reflect LLM errors or discoveries of new categories.
We also used human-defined taxonomies as references, acknowledging they might be inconsistent or incomplete.

\noindent \textbf{Internal Validity.}
Results depended on the stochastic behavior of DeepSeek-V3; we averaged ten runs to reduce variance. 
Outcomes may still vary with prompt design and batch size.

\noindent \textbf{External Validity.}
Our evaluation used SATD datasets from three domains.
While these covered diverse domains, they may not generalize to all types of software projects.

\vspace{-1mm}
\section{Conclusion} \label{conclusion}
\vspace{-1mm}

We investigated the extent to which LLMs could generate SATD taxonomies as an initial step toward automated open coding. Across three datasets, our LLM‑driven pipeline (i) recovered domain‑specific categories reported in prior studies (e.g., \textit{Layer Configuration} in ML), (ii) outperformed the naive baseline by as much as +0.13 F1 in comment assignment, and (iii) completed taxonomy construction for 448 comments in under two hours at a cost below \$1. Although full automation remains challenging, LLMs can reduce human effort. 
Future work will explore human‑in‑the‑loop pipelines and extend this approach to other open coding tasks beyond SATD taxonomy construction in software engineering.

\vspace{-1mm}
\section*{Acknowledgment}
\vspace{-1mm}
We gratefully acknowledge the financial support of: (1) JSPS for the KAKENHI grants (JP24K02921, JP25K03100, JP25K22845); (2) Japan Science and Technology Agency (JST) as part of Adopting Sustainable Partnerships for Innovative Research Ecosystem (ASPIRE), Grant Number JPMJAP2415, and (3) the Inamori Research Institute for Science for supporting Yasutaka Kamei via the InaRIS Fellowship.

\vspace{-1mm}
\bibliographystyle{IEEEtran}
\bibliography{references}
\vspace{-1mm}

\end{document}